\documentclass[a4paper,fleqn,usenatbib]{mnras}

\usepackage[T1]{fontenc}
\usepackage{ae,aecompl}

\usepackage{float}

\usepackage{graphicx}	
\usepackage{amsmath}	
\usepackage{amssymb}	
\usepackage{lineno}
\usepackage{afterpage}

\usepackage{enumitem,kantlipsum}
\setcitestyle{notesep={}}
\usepackage{newtxtext,newtxmath}
\title[cE Formation]{The formation pathways of compact elliptical galaxies }

\author[S. Deeley et al.]{
Simon Deeley$^{1}$\thanks{E-mail: s.deeley@uq.edu.au},
Michael J. Drinkwater$^{1}$, Sarah M. Sweet$^{1,2,3}$, Kenji Bekki$^{4}$, Warrick J. Couch$^{2}$, \newauthor Duncan A. Forbes$^{2}$\\
${}^1$School of Mathematics and Physics, University of Queensland, Brisbane, Queensland 4072, Australia\\
${}^2$Centre for Astrophysics \& Supercomputing, Swinburne University, Hawthorn, VIC 3122, Australia \\
${}^3$ARC centre of Excellence for All Sky Astrophysics in 3 Dimensions (ASTRO 3D)\\
${}^4$International centre for Radio Astronomy Research, The University of Western Australia, 35 Stirling Highway, Crawley, Western Australia, 6009, Australia \\
}

\date{Accepted XXX. Received YYY; in original form ZZZ}

\pubyear{2019}

\begin{document}
\label{firstpage}
\pagerange{\pageref{firstpage}--\pageref{lastpage}}
\maketitle
\begin{abstract}

Compact elliptical (cE) galaxies remain an elusively difficult galaxy class to study. Recent observations have suggested that isolated and host-associated cEs have different formation pathways, while simulation studies have also shown different pathways can lead to a cE galaxy. However a solid link has not been established, and the relative contributions of each pathway in a cosmological context remains unknown. Here we combine a spatially-resolved observational sample of cEs taken from the SAMI galaxy survey with a matched sample of galaxies within the IllustrisTNG cosmological simulation to establish an overall picture of how these galaxies form. The observed cEs located near a host galaxy appear redder, smaller and older than isolated cEs, supporting previous evidence for multiple formation pathways. Tracing the simulated cEs back through time, we find two main formation pathways; $32\pm5$ percent formed via the stripping of a spiral galaxy by a larger host galaxy, while $68\pm4$ percent formed through a gradual build-up of stellar mass in isolated environments. We confirm that cEs in different environments do indeed form via different pathways, with all isolated cEs in our sample having formed via in-situ formation (i.e. none were ejected from a previous host), and $77\pm6$ percent of host-associated cEs having formed via tidal stripping. Separating them by their formation pathway, we are able to reproduce the observed differences between isolated and host-associated cEs, showing that these differences can be fully explained by the different formation pathways dominating in each environment. 

\end{abstract}





\section{Introduction}
\label{introduction}

\begin{figure*}
\includegraphics[width=2\columnwidth]{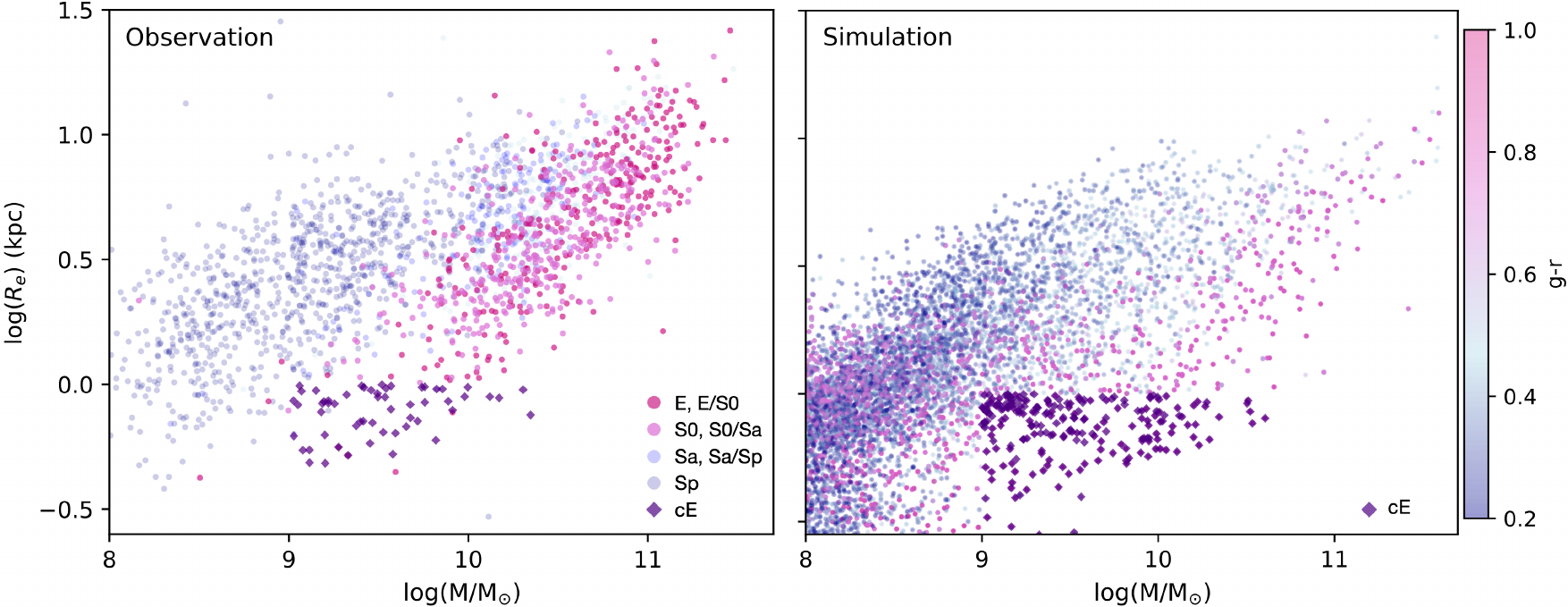}
   \caption{Selection of cEs in the radius-mass plane for the observed (left panel) and simulation (right panel) sample. The points in the left panel are coloured by their visual classifications, while in the right panel the points are coloured by their $g-r$ colour index, with an index of 0.6 roughly separating star forming and quiescent galaxies. Selected galaxies in our cE samples (with ${\rm R_{e} < 1 kpc}$) are highlighted by the indigo diamonds. The selected galaxies in the observed sample, most of which which were originally classified as 'unknown', form a low-mass extension of the main elliptical galaxy sequence.}
 \label{sample}
 \end{figure*}

Compact elliptical galaxies (cEs) are very small, high-mass galaxies which until recently have been difficult to identify and study owing to their rarity and small size. M32, a cE located in the vicinity of the Andromeda galaxy, was the first discovered and serves as the archetypal example for this class \citep{1973ApJ...179..731F}. They are typically defined to have a radius less than 1 kpc and a mass above $10^{8-9}$ solar masses, however there are variations in the exact definition used. Initially, cEs appeared to be a distinct class well separated from ultra-compact dwarf galaxies, which have typical radii below 200 parsecs and masses up to $10^{8} M_{\odot}$ \citep{2003Natur.423..519D,2017A&A...608A..53J}, as well as from the 'normal' elliptical galaxies and the less-compact dwarf elliptical galaxies \citep{2011MNRAS.414.3699M}. As searches and instrument sensitivity have improved and more compact galaxies are found, it is becoming evident that there is likely more of a continuum in systems from ultra-compact dwarfs (and even large star clusters) up to high-mass elliptical galaxies \citep{2011MNRAS.414.3699M,2014MNRAS.443.1151N} and the exact delineations between galaxy classes may be slightly arbitrary. For reasons outlined later, here we adopt a lower mass limit of $10^{9}M_{\odot}$ for our cE samples.

Many cEs appear to be offset from the elliptical galaxies in the mass-metallicity relation \citep{2018MNRAS.473.1819F}. They are observed to have smaller radii and higher metallicities \citep{2009Sci...326.1379C} than expected by these relations, suggesting that they are not a simple continuation of the elliptical galaxy population down to lower masses. This indicates that these galaxies are not forming through hierarchical growth like the rest of the elliptical population but are instead a different class of galaxy with a different formation pathway. The properties of cEs have been noted as being consistent with the observed population of high-redshift ($z\sim2$) compact passive galaxies, suggesting they could be relics of this earlier population \citep{2017MNRAS.468.4216Y,2011ApJ...739L..44D}. Recent studies have also suggested that at least some of these galaxies may go on to form the core of larger lenticular (S0) galaxies and other disk galaxies \citep{2018MNRAS.477.2030D,2015ApJ...804...32G}.

Most cEs have been found near much larger galaxies, and are likely physically associated with these galaxies, i.e.\ either interacting with them or orbiting them \citep{2017ApJ...835L...2Z}. However, an increasing number of cEs have also been identified in isolation, far away from any potential host galaxy \citep{2013MNRAS.430.1956H}. This finding has complicated the picture of how cEs form, since the formation mechanism needs to account for cEs in extremely different environments. \citet{2015Sci...348..418C} showed that significant numbers of compact ellipticals within cluster environments are capable of being ejected from the cluster entirely, which potentially explains why some cEs are in isolation. However, recent observational studies have shown that there are significant differences in the properties of host-associated and isolated cEs \citep{2020ApJ...903...65K,2018MNRAS.473.1819F}. \citet{2020ApJ...903...65K} showed that isolated cEs tend to follow the mass-metallicity relation of the larger ellipticals, suggesting that they are an extension of the normal elliptical galaxy sequence, while those near a large galaxy have higher metallicities than expected for their stellar masses. This indicates that rather than being ejected from clusters, isolated cEs may have formed in place via a completely different mechanism than the cluster cEs.

There are a number of different formation pathways which have been suggested for the formation of compact ellipticals. Simulations have shown that cEs can result from the tidal stripping of a disk galaxy through interactions with a larger host galaxy \citep{2001ApJ...557L..39B}, which would leave behind a compact core with a high metallicity. This pathway is supported by the observations of tidal streams around cEs located next to larger host galaxies \citep{2011MNRAS.414.3557H,2014ApJ...796L..14P}. Alternatively, cEs may form directly from a collapsing gas cloud or a hierarchical merging process, making up the lower-mass, lower-luminosity end of the elliptical galaxy sequence. For example, \citet{2015MNRAS.450.2327Z} showed that galaxies which initially form as gas-rich disks can undergo a compaction phase resulting in a compact high-mass galaxy, and \citet{2019ApJ...875...58D} showed that an initially gas-rich dwarf galaxy may be compacted by the combination of ram pressure stripping and tidal interactions with a large host galaxy. Such galaxies would be expected to feature lower metallicities than those formed from stripping, potentially explaining the properties of observed isolated cEs, but which is inconsistent with the host-associated cEs. Such direct formation may also contribute to the cluster sample \citep{2017MNRAS.470.4015M}. Simulations by \citep{2019MNRAS.489.2746U} have suggested that galaxies with the physical properties of cEs may form through the merging of star clusters located around the larger host galaxy. Signs of multiple pathways occurring have also been seen in kinematic measurements of cEs, with \citet{2021MNRAS.503.5455F} finding 5 cEs consistent with a stripped origin and 1 cE consistent with having formed intrinsically as a compact galaxy. \citet{2015ApJ...804...70G} measured the kinematics of 8 compact early-type galaxies in the Virgo cluster and found a range in kinematic structures showing signs of extensive interactions with the cluster environment, which may have lead to their compactness.

Intriguingly, hints of embedded structure have been seen in dwarf early-type galaxies \citep{2021A&A...647A.100S}, including spiral structure \citep{2000A&A...358..845J, 2021ApJ...912..149S}. While these galaxies are typically more extended than the cE population, this raises the possibility of embedded structures within the cEs;  observations of most cEs don't have the resolution needed to detect any internal spiral or disk structure, so whether they contain embedded structures is unexplored. Such internal structure, if present, may help in answering the question of how they formed. 

In this work we aim to determine which cE formation pathways are occurring in a cosmological context, and whether or not different formation pathways can explain the observed differences between isolated and host-associated cEs. In Section 2 we outline the observational and simulated data sample. In Section 3 we provide an overview of the methods used to derive the kinematics of the observed sample and to follow the evolutionary history of simulated galaxies. In Section 4 we present the observed comparison between isolated and host associated cEs, and identify different formation pathways in the simulation, and finally our results are discussed in Section 5. Our observational results assume a ${\rm \Lambda CDM}$ cosmology with ${\rm \Omega_{m}}=0.3$, ${\rm \Omega_{\lambda}}=0.7$ and ${\rm H_{0}=70 km s^{-1} kpc^{-1}}$. The simulation used in this study assumes a ${\rm \Lambda CDM}$ cosmology with ${\rm \Omega_{m}}=0.3089$, ${\rm \Omega_{\Lambda}}=0.6911$ and ${h=0.6774}$.

\begin{figure*}
\includegraphics[width=2\columnwidth]{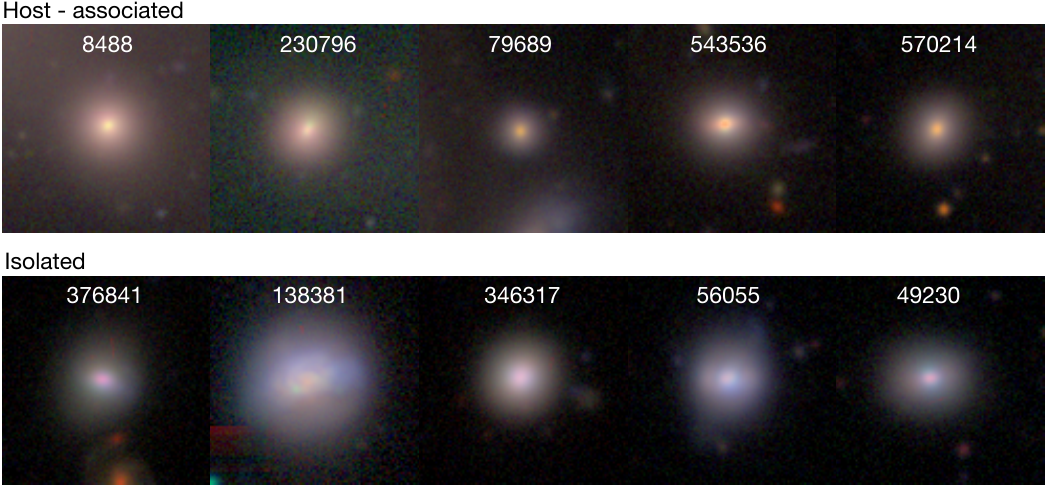}
   \caption{Optical HSC imaging of example host-associated (top row) and isolated (bottom row) compact elliptical galaxies. The isolated cEs tend to have a bluer appearance and two of them (56055 and 138381) show interesting features hinting at an embedded structure. }
 \label{images}
 \end{figure*}

\section{Data}
\label{data}

\begin{table*}

\caption{Observational sample of galaxies with stellar masses greater than $10^{9}M_{\odot}$ and effective radius $R_{e}$ < 1 kpc.}
\centering
\begin{tabular}{ccccclccccc}
\cline{1-5}
\cline{7-11}

   SAMI ID &      RA &    DEC &   log Mass (M$_{\odot})$ &   Re (kpc) &             &   SAMI ID &      RA &    DEC &   log Mass (M$_{\odot})$ &   Re (kpc) \\
\cline{1-5}
\cline{7-11}
      8450 & 182.421 &  0.658 &                    9.4 &       0.71 &     &    298593 & 220.601 & 28.658 &                    9.5 &       0.92 \\
      8488 & 182.629 &  0.673 &                    9.6 &       0.64 &     &    321099 & 221.197 & 28.673 &                    9.7 &       0.71 \\
      9061 & 184.824 &  0.824 &                    9.5 &       0.96 &     &    346317 & 133.27  & 28.824 &                    9.7 &       0.73 \\
     16392 & 218.528 &  0.633 &                    9.2 &       0.63 &     &    372385 & 135.561 & 28.633 &                    9.1 &       0.49 \\
     16515 & 219.197 &  0.718 &                    9.6 &       0.53 &     &    372446 & 135.886 & 28.718 &                   10.2 &       0.88 \\
     30467 & 175.393 & -1.2   &                   10.3 &       0.75 &     &    375598 & 129.906 & 26.8   &                    9.4 &       0.99 \\
     40765 & 182.897 & -0.7   &                    9   &       0.62 &     &    376841 & 135.42  & 27.3   &                    9.5 &       0.89 \\
     41319 & 185.436 & -0.837 &                    9.1 &       0.54 &     &    381208 & 131.684 & 27.163 &                    9.5 &       0.58 \\
     49230 & 220.176 & -0.63  &                    9.2 &       0.86 &     &    381217 & 131.685 & 27.37  &                    9.6 &       0.99 \\
     56055 & 184.377 & -0.232 &                    9.2 &       0.77 &     &    402988 & 178.355 & 27.768 &                    9.5 &       0.95 \\
     69462 & 174.874 &  0.117 &                    9.2 &       0.84 &     &    485837 & 217.625 & 28.117 &                    9.5 &       0.64 \\
     79689 & 222.713 &  0.095 &                    9.1 &       0.55 &     &    486003 & 218.125 & 28.095 &                    9.5 &       0.7  \\
     79706 & 222.703 &  0.094 &                    9.3 &       0.56 &     &    486139 & 218.607 & 28.094 &                    9.6 &       0.79 \\
     91568 & 212.759 &  0.609 &                    9.9 &       0.79 &     &    508006 & 215.427 & 28.609 &                   10   &       0.89 \\
     93816 & 222.518 &  0.555 &                    9   &       0.86 &     &    511923 & 216.693 & 28.555 &                    9.4 &       0.68 \\
    106618 & 216.811 &  0.99  &                    9.7 &       0.83 &     &    517590 & 132.943 & 28.99  &                    9.5 &       0.89 \\
    138291 & 181.858 & -1.669 &                    9.3 &       0.91 &     &    521348 & 129.065 & 26.331 &                    9.9 &       0.86 \\
    138381 & 182.069 & -1.782 &                    9.4 &       0.97 &     &    537367 & 184.961 & 26.218 &                    9.2 &       0.94 \\
    144682 & 181.035 & -1.417 &                    9.1 &       0.99 &     &    543536 & 211.896 & 26.583 &                    9.8 &       0.6  \\
    186202 & 183.478 & -1.615 &                    9.8 &       0.98 &     &    543895 & 213.272 & 26.385 &                    9.8 &       0.96 \\
    205155 & 140.661 & -0.384 &                    9.7 &       0.97 &     &    551194 & 139.396 & 27.616 &                    9.8 &       0.85 \\
    228767 & 218.631 &  1.188 &                    9.2 &       0.85 &     &    570214 & 222.742 & 29.188 &                    9.4 &       0.71 \\
    230789 & 181.072 &  1.917 &                    9.3 &       0.52 &     &    574386 & 135.624 & 29.917 &                    9.5 &       1    \\
    230796 & 181.153 &  1.893 &                    9.2 &       0.48 &     &    600446 & 135.55  & 29.893 &                   10.1 &       0.89 \\
    238218 & 213.647 &  1.752 &                   10   &       0.97 &     &    617665 & 212.781 & 29.752 &                    9.1 &       0.94 \\
    238411 & 214.247 &  1.574 &                    9.7 &       0.68 &     &    619905 & 222.17  & 29.574 &                   10.3 &       0.95 \\
    278771 & 133.613 &  0.921 &                    9.1 &       0.83 &     &    621917 & 129.036 & 28.921 &                    9.2 &       0.84 \\
    297138 & 214.754 &  1.412 &                    9.9 &       0.96 &     &    719329 & 131.478 & 29.412 &                    9.3 &       0.84 \\\cline{1-5}
\cline{7-11}
\end{tabular}
\label{table}
\end{table*}

\subsection{Observational Sample}

The observational data products used in this work were created by the SAMI Galaxy Survey \citep{2015MNRAS.447.2857B}. The Sydney-AAO Multi-object Integral field spectrograph \citep[SAMI;][]{2012MNRAS.421..872C} was mounted at the prime focus of the 3.9m the Anglo-Australian Telescope, which provided a 1 degree diameter field of view. SAMI used 13 fused fibre bundles \citep[hexabundles;][]{2011OExpr..19.2649B, 2014MNRAS.438..869B} with a high (75 percent) fill factor. Each bundle contains 61 fibres of 1.6 arcsec diameter resulting in each hexabundle having a diameter of 15 arcsec. The hexabundles, as well as 26 sky fibres, were plugged into pre-drilled plates using magnetic connectors. SAMI fibres were fed to the double-beam AAOmega spectrograph \citep{2006SPIE.6269E..0GS}. For the SAMI Galaxy Survey its 570V grating was used with the blue arm (3700-5700A), giving a resolution of R=1730 (${\rm \sigma=74 km/s}$), and the R1000 grating with the red arm (6250-7350A) giving a resolution of R=4500 (${\rm \sigma=29 km/s}$) \citep{2017ApJ...835..104V}. At least six pointings on each galaxy were weighted and combined to produce data cubes with a pixel scale of $0.5\times0.5$ arcseconds \citep{2015MNRAS.446.1567A, 2015MNRAS.446.1551S}. Here we use data products released in Data Release 3 \citep{2021MNRAS.505..991C}. The effective radius for each galaxy was derived from a single component fit of SDSS imagery using the code GALFIT3 \citep{2010AJ....139.2097P, 2011MNRAS.413..971D}.The stellar masses were derived from mass/light ratios based on $g-i$ colours \citep{2015MNRAS.447.2857B} and the global star formation rates were derived from the galaxy's H$\alpha$ flux \citep{2018MNRAS.475.5194M}.

The left panel of Figure~\ref{sample} displays the entire SAMI galaxy sample coloured by visual morphology \citep{2016MNRAS.463..170C}. We selected galaxies with stellar masses greater than $10^{9} M_{\odot}$ and effective radii less than 1 kpc to form our observational cE sample. Our sample forms an extension of the early-type galaxy sequence to smaller masses and radii. We were unable to include lower mass compact galaxies because the SAMI survey includes very few such galaxies \citep{2015MNRAS.447.2857B}, and those observed were classed as late-type spiral galaxies (see Figure~\ref{sample}). We identified 27 galaxies which matched these criteria and for which we have IFU observations. An additional 29 cEs were identified in the SAMI input catalogue which weren't targeted during the survey and hence don't have IFU data; we include these additional galaxies in the analysis of global properties. The catalogue ID numbers, locations, masses and radii for all of these galaxies are shown in Table~\ref{table}. The full SAMI sample had been visually classified using SDSS imaging, however due to the small size of these compact galaxies, nearly all galaxies in our sample were classified as 'unknown'. The location of these galaxies in the size-mass plane in the SAMI sample is also shown in the left panel of Figure~\ref{sample}.

To confirm their nature as compact elliptical galaxies, we visually inspected higher-resolution imagery from the Hyper Suprime Cam (HSC) \citep{2019PASJ...71..114A}. Based on these images, two galaxies were removed from the sample, with one found to be a clumpy dwarf galaxy and the other an extended disk galaxy with a very bright core. All remaining galaxies have a very compact, spherical and centrally concentrated structure typical of this class; examples are presented in Figure~\ref{images}. 

\subsection{Simulation Sample}

For the simulation component of this study we use the IllustrisTNG-50 cosmological simulation \citep{2019ComAC...6....2N,2019MNRAS.490.3196P,2018MNRAS.475..624N,2018MNRAS.475..676S,2018MNRAS.477.1206N,2018MNRAS.480.5113M}. The high resolution of this simulation (with each baryon particle representing $8.5 \times 10^{4} ~\rm M_{\odot}$) allows us to resolve these small compact galaxies in higher resolution and confidently identify them as true galaxies rather than spurious fragments. However, we note that due to the small volume of this simulation, we miss the highest-mass cluster environments; the largest system within this simulation is similar to the Virgo cluster with a total mass of around $10^{14} M_{\odot}$.   

To identify compact elliptical galaxies in the IllustrisTNG-50 simulation, we applied the same selection criteria as used for the observational sample - selecting all galaxies with stellar masses above $10^{9}  M_{\odot}$ and effective radii less than 1 kpc. We found a total of 274 galaxies matching these criteria. We removed all galaxies which only came into existence late in the simulation, as the majority of these are spurious galaxies where, for example, a fragment of a larger galaxy has been separated into its own halo by the halo finding algorithm. This leaves us with a total of 226 compact galaxies. The resulting sample is shown in the mass-radius plane in Figure~\ref{sample}, along with the other galaxies in the simulation (coloured by their colour index). 

We defined an additional simulation sample to include the lower-mass compact galaxies and bridge the gap with ultra-compact dwarf (UCD) galaxies. We selected galaxies with ${\rm R_{e} < 1 kpc}$ and stellar masses in the range $10^{8} < M_{\odot} < 10^{9}$. Over 2,500 galaxies meeting these criteria were found in the simulation; we randomly selected 500 of these galaxies and, after analysing their formation histories and again removing those which formed late in the simulation, we were left with 435 galaxies. The main results in this paper focus only on the complete higher-mass sample; we include results for this lower-mass sample in the appendix. UCD galaxies with ${\rm R_{e} < 200 pc}$ and  $ M_{\odot} < 10^{8}$ are very rare in the simulation, likely because the simulation's resolution makes it difficult for such galaxies to remain tightly bound, particularly within highly dynamical group environments. Therefore we are unable to include these galaxies using the methods we employ here; see \citet{2019lgei.confE..33M} for an analysis of UCDs in a cosmological simulation.

In order to investigate the evolutionary histories and identify merger and infall events for our compact galaxies, we use the Sublink merger trees created for IllustrisTNG-50 \citep{2015MNRAS.449...49R}. The colour indices are derived from the stellar photometries created by \citet{2018MNRAS.475..624N}.

\section{Methods}
\label{methods}

\begin{figure}
\includegraphics[width=1\columnwidth]{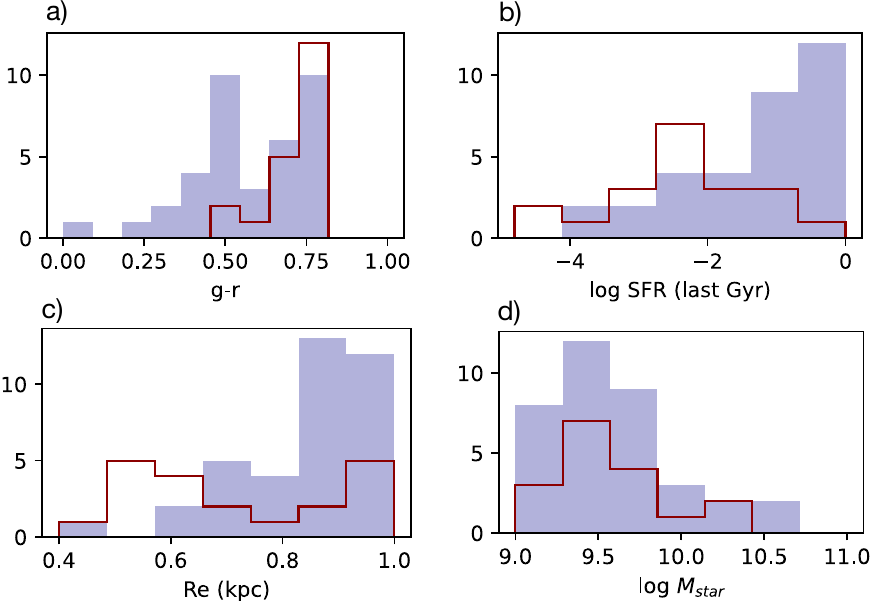}
   \caption{Colour, star formation rate, effective radii and stellar mass distributions for isolated (blue filled histograms) and host-associated (red empty histograms) observed cEs. Host-associated cEs are redder, have lower star formation rates and are more compact than their isolated counterparts on average, suggesting that they make up a seperate population with a different formation path. Meanwhile, they have a similar mass distribution, showing that the observed differences are not due to mass differences alone.}
 \label{distribution}
 \end{figure}

\begin{figure*}
\includegraphics[width=1.8\columnwidth]{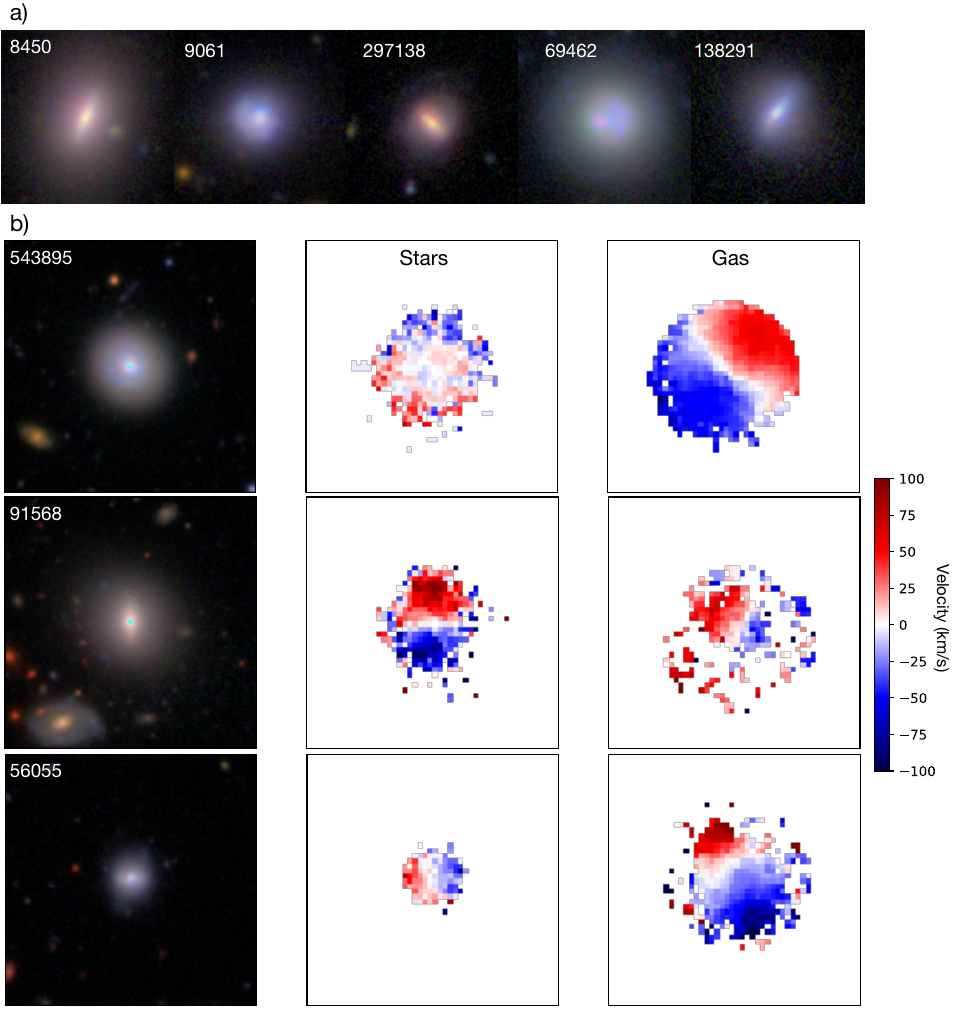}
   \caption{Compact galaxies with internal structure and misaligned kinematics. The top row shows galaxies with embedded structures which may be disks viewed from edge on or face-on, or internal regions of ongoing star formation. b) shows three isolated galaxies which have misaligned star and gas kinematics. }
 \label{disks}
 \end{figure*}

\subsection{Associating cE galaxies with potential hosts}

Identifying whether two galaxies appearing close together on the sky are physically associated with each other can be difficult. The separation along the line-of-sight is determined from the differences in redshift between the two galaxies using the redshift-distance relation. However this is complicated by the changes in redshift caused by the peculiar motions of galaxies through space. In particular, when looking at a small galaxy in a tight orbit around a large host, the high orbital velocity (resulting in a large difference in the line-of-sight velocities to the observer) can result in a large difference in the redshift of the galaxy relative to its host (up to 1 Mpc when translated into a distance assuming standard cosmology), meaning that two tightly associated galaxies may be seen to have large separations in redshift space. 

Therefore, to mitigate these effects and increase our certainty in any physical associations, we take advantage of the Galaxy and Mass Assembly (GAMA) group catalogue \citep{2011MNRAS.416.2640R}, which contains our entire observational sample. The catalogue was constructed using a friends-of-friends algorithm, taking into account both the projected on-sky separation and the separation in redshift space between galaxies. To identify which of our cE galaxies is associated with larger hosts, we firstly find the cEs which have been assigned to a galaxy group, and hence identify all other galaxies which are likely physically associated with those cEs. Within that galaxy group, we then take all potential host galaxies with stellar masses greater than $10^{10.5}M_{\odot}$ and identify which one of these is the closest (in projection) to the cE galaxy in question (we note that the results presented here do not change significantly when this mass selection is varied between $10^{10}$ and $10^{11}M_{\odot}$). The angular distance to this galaxy, translated to a spatial separation assuming standard cosmology and the redshift of the cE galaxy, is then assigned as the distance to the nearest potential host. 20 cE galaxies were found to reside within 500 kpc of a high-mass host galaxy. For the remaining cE galaxies not located within a group (or located within a group containing no high-mass galaxies), we know they are not physically associated with any high-mass galaxies and so we don't need to worry about missing close hosts due to tight high-velocity orbits. We therefore perform a 3D search in projection-redshift space for the nearest  >$10^{10.5}M_{\odot}$ galaxy and again use the separation to this galaxy as the distance to the nearest high-mass galaxy.

\subsection{Kinematic profiles}

For the galaxies in our sample which have been observed by SAMI, we derived radial kinematic profiles from the 2D kinematic maps for each galaxy, and investigated the degree of rotational support. For the gas and stellar velocity maps, spaxels with uncertainties greater than $\pm15$ km/s were removed from the analysis. For the velocity dispersion maps, following \citet{2017ApJ...835..104V} we removed spaxels where the uncertainty is greater than 10 percent of the dispersion value plus 25 km/s. From the remaining spaxels, we constructed the radial profile of the velocity ($v$) using the maximum velocity in each radial annuli, and radial profiles of the velocity dispersion ($\sigma$) using the median values in each bin. The degree of rotational support, measured using the ratio $v/\sigma$, was then calculated by:

\begin{equation}
\left( \frac{v}{\sigma} \right)^{2} = \frac{\sum_{i=0}^{N_{spx}}{F_{i}{V_{i}^{2}  }}} {\sum_{i=0}^{N_{spx}}{F_{i}\sigma_{i}^{2}}},
\label{refinement_eq}	
\end{equation}

where $F_{i}, V_{i}$ and $\sigma_{i}$ are the flux, velocity and velocity dispersion of the $i$th spaxel respectively \citep{2018NatAs...2..483V}. 

For the simulation sample, we use the approach detailed by \citet{2021MNRAS.508..895D}  \citep[following the method outlined in][]{2019MNRAS.487.2354B} to create kinematic maps equivalent to the observations in SAMI. Briefly, we place the SAMI observational window over the galaxy in the x-y plane, scaled such that 3 effective radii are contained within SAMI's field-of-view. For each spaxel in the SAMI window, we created a histogram of the velocities of all stars in the galaxy, with each star weighted according to its distance to that spaxel by a Gaussian centred on that spaxel with a full-width-half-maximum of 2.03", representing SAMI's average point spread function. The peak velocity of the resulting histogram is then taken to be the line-of-sight velocity measured in that spaxel, and the histogram's standard deviation is used for the velocity dispersion. We then followed the same approach as above for the radial profiles and $v/\sigma$ calculation.

\subsection{Tracing histories in IllustrisTNG-50}

In order to investigate the evolutionary history of our simulation sample, we follow the same process as that used by \citet{2021MNRAS.508..895D}. To identify merger events, we use the LHaloTrees catalogue \citep{2005Natur.435..629S} and follow the galaxy's progenitors backwards through time. To find merger events, we look for a new secondary progenitor appearing along side the primary galaxy (i.e. when the 'Next Progenitor' flag is non-zero), which must have merged into the primary galaxy before the following snapshot. This following snapshot is then used as the time of the merger.

For this work, we used the stellar masses of the primary and secondary halo to calculate their mass ratios, since this can be more directly linked to observations. A well-known issue in merger trees is the flow of particles from the secondary to the primary halo before the merger occurs; if the merger ratio is calculated at the point of merger, the ratio can be significantly underestimated. We therefore trace both the primary halo and the secondary halo back 20 snapshots (or up to the beginning of their existence, if this is less than 20 snapshots) and identify the point at which the secondary is at its greatest mass. The mass of the primary and secondary halo at this snapshot is then used to calculate the merger ratio. In addition, to counteract the known halo switching problem (where close halos can be swapped around and incorrectly assigned to the primary or secondary galaxy, see for example \citet{2017MNRAS.472.3659P}) and keep the merger ratio below 1, we looked for switches in the primary and secondary halo masses during the merger and swapped the masses around when this occurred. The merger event is flagged as a minor merger if the stellar mass ratio of the merging galaxy and the main progenitor is between 1:10 and 1:3, or a major merger if the mass ratio is greater than 1:3.

To identify infall events, where a galaxy falls into a significantly larger group or cluster, we tracked the total mass of the host group halo for each galaxy through time and flagged any events where this mass suddenly increased by half an order of magnitude. Such an increase occurs between subsequent snapshots when the friends-of-friends group finding algorithm assigns the galaxy to the larger group in the following snapshot.

\section{Results}

Firstly, we present the results of the observational study, comparing the properties of isolated and host-associated cEs. We then use the simulation sample to identify three different formation pathways, and finally show that these different pathways lead to the observed differences between isolated and host-associated cE galaxies.

\subsection{Observations}

\begin{figure*}
\includegraphics[width=2\columnwidth]{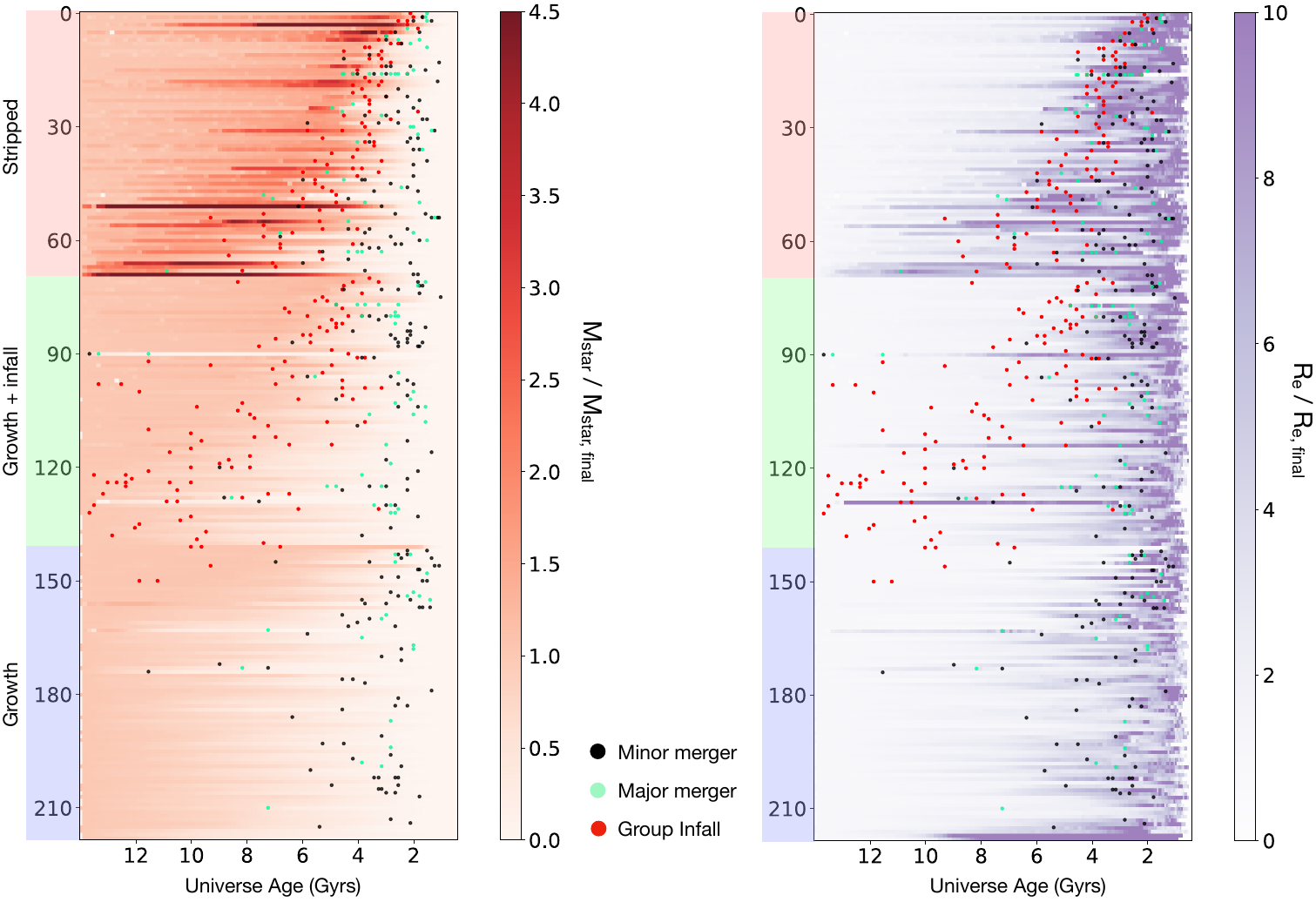}
   \caption{Evolution of the galaxy mass (left) and effective radius (right) relative to their final present-day values for the entire cE sample. Galaxies are stacked along the y-axis, with each line showing the evolution of a single galaxy. Red points correspond to cluster infall, green points to major mergers and black points to minor mergers. In the 'stripped' group, galaxies firstly increase then decrease their mass after the infall event, while those in the 'growth' and 'growth+infall' groups steadily grow their mass. In the evolution of galaxy radius, those in the stripped group initially building up to a much larger radius before shrinking, also indicating stellar stripping. The galaxies in the two lower groups also initially grow to larger radii before shrinking again, but to a lesser extent. In this case, this may be due to ongoing star formation (reflected in the increasing mass in the left-hand plot) becoming concentrated in the centre.}
 \label{hist_mass}
 \end{figure*}

\subsubsection{Physical Properties}

Figure~\ref{distribution} displays the colour, star formation rate, effective radii and stellar mass distributions of the observed cE sample, divided into isolated (blue) and host-associated (red) cEs. Isolated cEs are generally bluer, however there is also a significant red population present, leading to a binomial colour distribution. In contrast, all host-associated cEs fall into a single red population, with the majority having a colour index redder than 0.6. The isolated galaxies also peak towards higher rates of recent star formation, reflective of their blue colour. They also trend towards higher effective radii, while the host-associated population has a flatter distribution. Lastly, the stellar mass distributions are the same for both samples, showing that the other observed differences are not due to a difference in mass. These differences are in agreement with previous work \citet{2020ApJ...903...65K}, and further suggests cEs in different environments are forming through different pathways.

\subsubsection{Morphology and Embedded Disks}

Many of the isolated galaxies show evidence of some interesting substructures. Some of these, such as 376841 and 49230 feature blue cores, suggesting ongoing star formation in the very centre of the galaxy. These regions could be creating a central build-up of the galaxy's stellar mass, leading to the compact nature of these galaxies. 

More surprisingly, many feature signs of disk-like structures - examples of such galaxies are presented in Figure ~\ref{disks}. Edge-on embedded disks are also seen in redder cE galaxies, however we note that these may be a central bar rather than a disk. In a wider parameter search of the GAMA survey for galaxies meeting the size and mass criteria of a compact elliptical, 15 red ( $g-r > 0.6$) and 27 blue compact galaxies were seen to have such an embedded disk-like structure, suggesting that while not occurring in the majority of cases, there are a significant number of compact galaxies which feature disk-like structures. Disks have previously been observed in dwarf elliptical galaxies \citep{2003A&A...400..119D}, however to our knowledge they have not been observed in such compact galaxies previously. Disk structures would not be expected in a cE resulting from the stripping of a larger galaxy, which suggests such galaxies may be forming through an alternative pathway. Alternatively, they could have formed after the stripping occurred due to the infall of fresh star-forming material. 

Figure ~\ref{disks} also shows the kinematic maps for the stellar and gas components for three of these galaxies with unusual kinematics. In these cases, the two components are misaligned, in one case by around 180 degrees and in the other cases by less than 90 degrees. This may indicate a gas-rich merger or accretion of star-forming material from the galaxy's surroundings, and the compact nature of these galaxies may make such misalignments long-lived \citep{2022MNRAS.517.2677R}.

\begin{figure*}
\includegraphics[width=2\columnwidth]{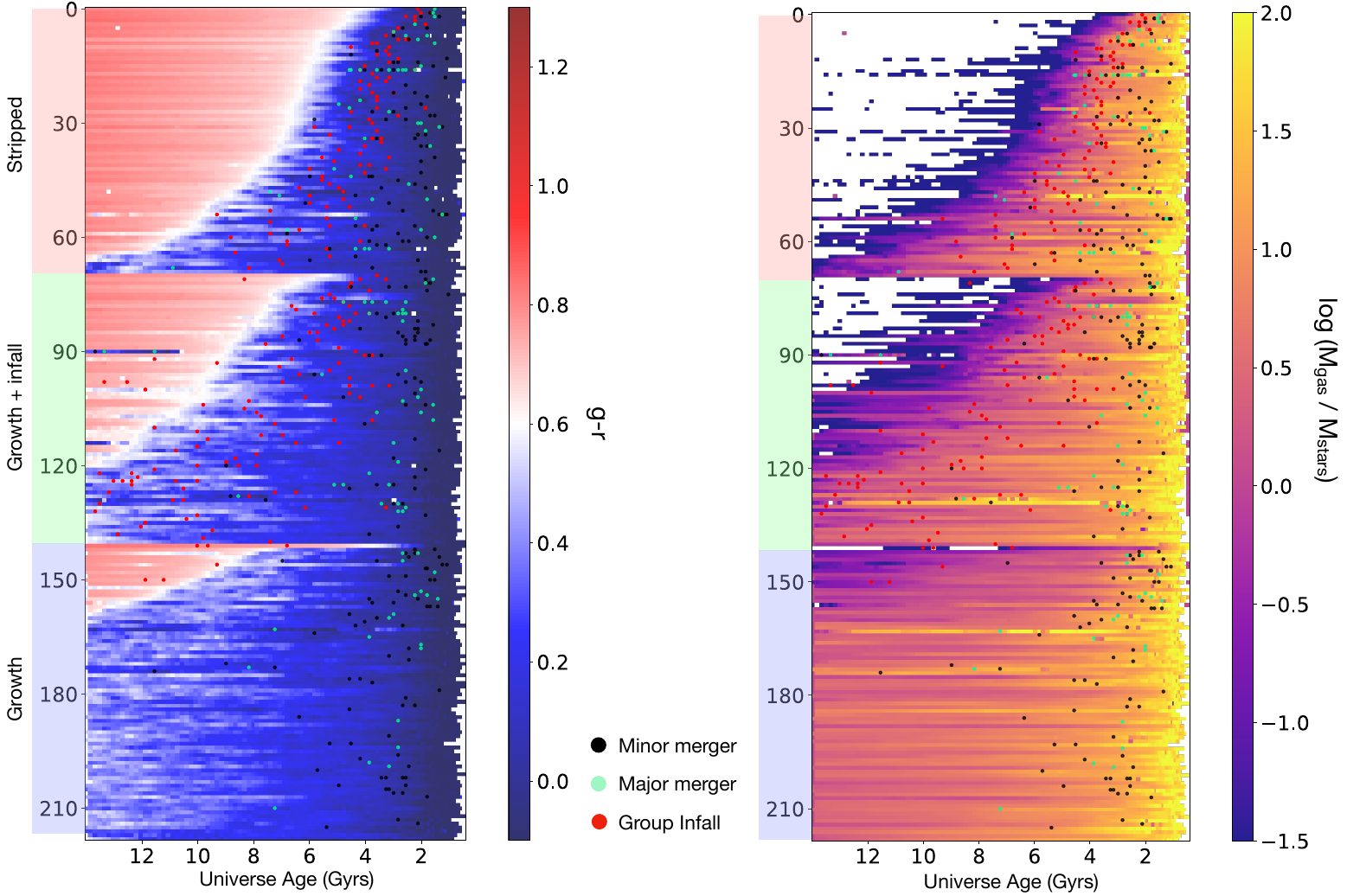}
   \caption{Colour (left) and total gas (right) histories of our cE sample, organised into the two mass-evolution pathways as identified in Figure 7. The galaxies in the stripped (top) and growth + infall (lower) groups have their gas stripped out after the infall event, bringing an end to their star formation and giving them a red colour. Those which remain in isolation retain large amounts of their gas and continue to form new stars up to the present day.}
 \label{hist_colour}
 \end{figure*}

\subsection{Identifying formation pathways in IllustrisTNG}

To test if the formation pathways leading to isolated and host-associated compact ellipticals are indeed different, we identified cEs within the IllustrisTNG-50 simulation and followed their formation back through time to determine the process through which they formed. We then used the identified pathways to predict the observed differences between cE populations in different environments.

\subsubsection{Mass evolution}

The two main proposed formation pathways involve either a large progenitor galaxy getting stripped by gravitational interactions leaving behind a dense core, or in-situ formation whereby the galaxy forms from a collapsing gas cloud in isolation and gradually builds up its mass (i.e. a continuation of the elliptical sequence down to lower masses). In the first case, the galaxy would be seen to grow to a large stellar mass before losing a significant fraction during the stripping process, while in the second case we would expect to see increasing stellar mass with time. Therefore we initially investigated the mass evolution of the cEs to look for these pathways. The sample was separated into those which at some time in their history reached a total stellar mass of at least 30 percent greater than their final mass (i.e. had experienced significant mass loss) and those which had not. 

The stellar mass evolution is displayed in the left panel of Figure~\ref{hist_mass}, where the mass at a given time is shown relative to the galaxy's final mass at the end of the simulation. In addition, major merger events, minor merger events and halo infall events are marked by green, black and red points respectively. Both types of mass evolution are present, with $32\pm6$ percent of cEs featuring a significant stripping of stellar mass after an initial rise and the remaining $68\pm4$ percent increasing in mass with time. In addition to the mass evolution divide, the mass growth group has been further separated into those which have fallen into a galaxy group and those which haven't, for reasons which will become clearer in the following section. For compact galaxies below $10^{9} M_{\odot}$, the relative numbers of galaxies in these groupings changes significantly; only $12\pm5$ percent experienced mass stripping while the remaining $88\pm3$ percent have a steady or increasing mass with time - see Figure~\ref{low_mass} in the Appendix.

In some cases the mass loss is very dramatic, with the galaxy at one stage being 5 times more massive than the final mass of the cE. Another important feature is the fact that the mass loss occurred quickly after the galaxy fell into the host halo of a significantly larger galaxy or galaxy cluster, providing an initial sign that this mass loss is indeed due to tidal stripping. The second and third groups feature a very gradual build-up of mass, continuing up to the final snapshot in many cases. This suggests that these galaxies are not forming through the merging of large star clusters as has been previously proposed (noting however that such galaxies forming from smaller star clusters may not be visible in Illustris due to the simulation's particle resolution). Instead, they appear to be either continually forming new stars or accreting stars from the surrounding environment.

\begin{figure*}
\includegraphics[width=2\columnwidth]{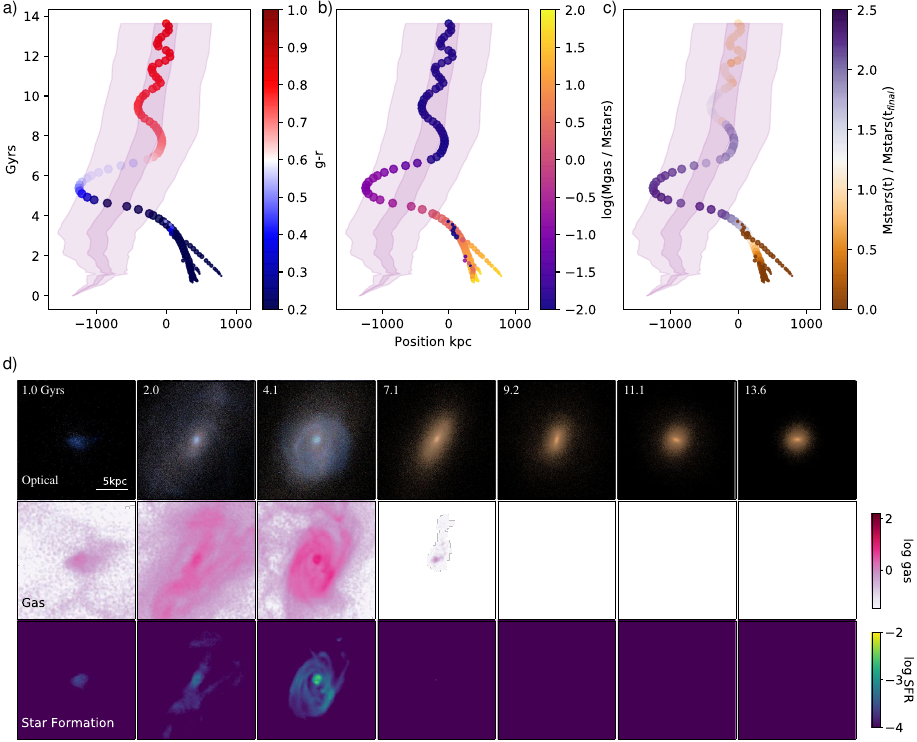}
   \caption{An example of a cE forming via the stripping process. The top row shows position-time plots of the galaxy's progenitor halos, including all those which have merged into the main progenitor. The diagram in a) is coloured by the galaxy's colour index, b) by the gas mass (normalised by the stellar mass), and c) by the stellar mass relative to the final stellar mass, illustrating the mass growth and subsequent stripping. The purple region shows the location of the final host halo traced back in time, with the lighter region showing the R200 radius and the darker region showing one quarter of this radius. d) shows the optical image created using the radiative transfer code SKIRT \citep{2015A&C.....9...20C}, along with the distribution of gas and star formation, at different stages of the galaxy's evolution. A large star-forming spiral enters a group environment, where it rapidly loses gas and enters a tight orbit around the central host galaxy. The outer stellar disk is gradually stripped away, leaving behind the dense core as a compact elliptical. }
 \label{stripped}
 \end{figure*}

\begin{figure*}
\includegraphics[width=2\columnwidth]{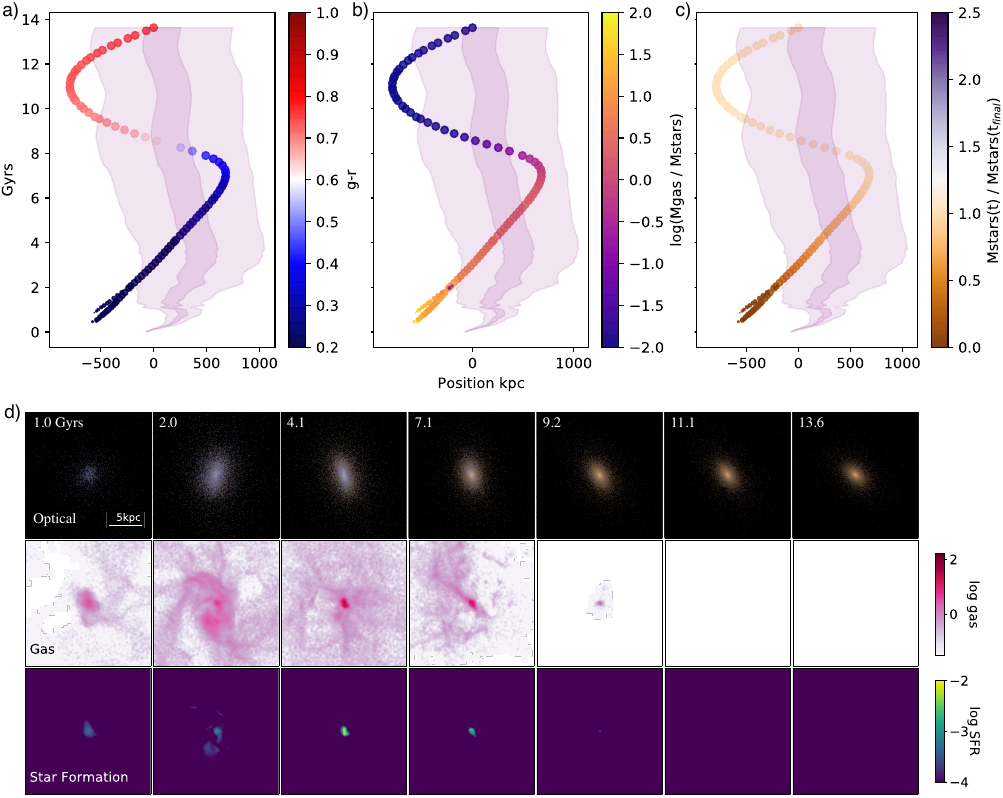}
   \caption{An example of a cE forming in-situ and then falling into a larger galaxy group, showing the same panels as those in Figure~\ref{stripped}. The galaxy is initially quite diffuse, then becomes concentrated as star formation shifts towards the centre. After falling into the group, the gas is quickly removed and the galaxy becomes passive after its first passage through the centre (note that the first cross over the centre of the host group in a), b) and c) is only in projection, i.e. the galaxy is still outside the group at this stage).}
 \label{infall}
 \end{figure*}

\begin{figure*}
\includegraphics[width=2\columnwidth]{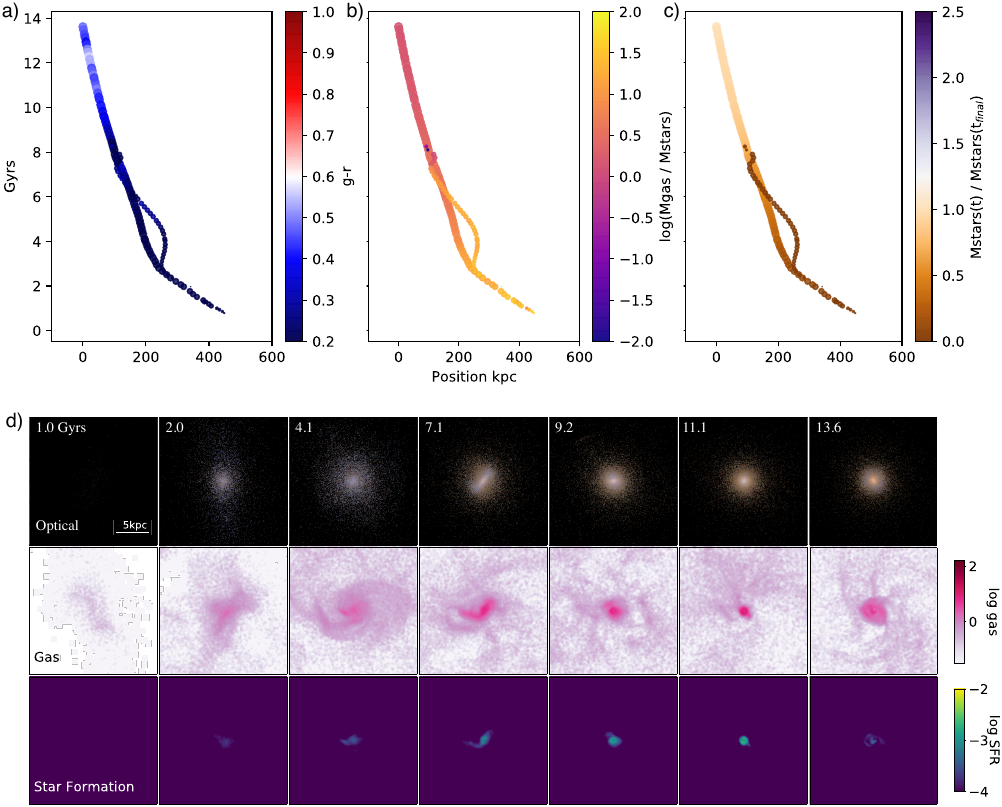}
   \caption{Example of a cE forming in-situ, shown in the same format as Figure 7. After initially having star formation activity through the galaxy, star formation becomes concentrated in the centre where it builds up a large core, resulting in a more concentrated galaxy.}
 \label{growth}
 \end{figure*}

\subsubsection{Radius evolution}

The evolution of the effective radius of the galaxies, relative to their final radius, is shown in the right-hand panel of Figure~\ref{hist_mass}. The galaxies which are stripped of a significant amount of their mass also experience a concurrent decrease in their radius, showing that the mass is stripped from the outskirts of the galaxy. Prior to the mass-stripping event, many of these galaxies were up to 10 times larger than their final compact form, indicative of an extended disk-like component. The cEs forming through the two growth pathways were also more extended in the past, and gradually became more compact towards the present day. However in this case this evolution is likely due to the continued build-up of new stars towards the centre of these galaxies; rather than losing mass from the outskirts, they are gaining mass in their centres, which makes their overall mass profiles more compact and has the effect of decreasing the radius within which half their mass is contained. 

\subsubsection{Colour and gas evolution}

Figure~\ref{hist_colour} displays the evolution of the galaxy's colours and gas content, with the galaxies being shown in the same arrangement as those in Figure~\ref{hist_mass}. The galaxies in the stripped and infall group are in all but two cases red in the final snapshot. The transition from a blue star-forming galaxy to a passive red galaxy occurs within around 2-3 billion years after the galaxy falls into the halo of the larger galaxy or cluster, coinciding with significant gas stripping and a sharp cutoff in star formation activity. This strongly suggests that the environment of the new host group has a dramatic impact on the galaxy and its ability to form stars. 

In contrast, the vast majority of those in the growth group remain blue right up to the present day, retaining a significant amount of their gas. This is accompanied by a small yet constant rate of star formation, showing that these galaxies are yet to run out of star-forming material.

\subsubsection{Stripped Pathway}

A randomly selected example of a compact elliptical forming via the stripping pathway is shown in Figure~\ref{stripped}. The galaxy initially develops into a typical large, blue star-forming spiral galaxy. This galaxy then falls into a large halo, rapidly losing its star-forming material via ram pressure stripping, and becomes a red passive spiral galaxy. As its orbit around the central galaxy gets tighter and as its outer stars begin to be stripped away, its spiral structure is lost and the galaxy transitions to an S0. At this stage, the galaxy has followed the same evolutionary pathway as an 'infall' S0 in \citet{2021MNRAS.508..895D}, except that it falls into a tighter orbit around the central galaxy and hence experiences greater tidal forces. As it continues to orbit the central galaxy, stars in the disk continue to be stripped away until all that is left is the central bulge. The final result is a passive, compact elliptical galaxy in a tight orbit around the central host galaxy.

\subsubsection{Growth + infall Pathway}

An example of a galaxy which experiences continual growth (in contrast to mass stripping) while also experiencing an infall into the vicinity of a larger galaxy or group is shown in Figure~\ref{infall}. Prior to the infall, these galaxies follow an evolutionary path very similar to that of the growth pathway, usually becoming compact objects before the infall itself. Once the galaxy falls into the denser environment, however, the galaxy loses most of its gas and hence the amount of star formation activity decreases. Eventually, star formation activity ceases and the galaxy becomes a compact passive galaxy with an elliptical morphology.

\subsubsection{Growth Pathway}

Figure~\ref{growth} displays a typical example of a compact galaxy forming via the 'growth' pathway. Initially, a gas cloud collapses to form a diffuse dwarf galaxy with star formation activity occurring throughout the galaxy. Gradually, the gas becomes more concentrated in the centre, until the majority of gas within the galaxy is located within the core. The region of star formation follows, gradually becoming more contracted towards the centre. The decreasing star formation in the galaxy's outskirts is reflected in the g-band luminosity images, where the galaxy appears to contract as the bright, short-lived stars in the outskirts disappear. The contracted star formation continues to build up stellar mass within the core, which has the result of gradually increasing the central concentration of the galaxy to the point where its half-mass radius falls below 1 kpc. The final stellar distribution of the galaxy takes the form of a compact elliptical, however interestingly there is a tight embedded disk (only visible in the g-band luminosity images) where star formation activity continues. The majority of these galaxies are located in relatively isolated environments; the nearest galaxy group to the galaxy shown in Figure~\ref{growth}  lies at a distance of around 1 Mpc.

\begin{figure*}
\includegraphics[width=2\columnwidth]{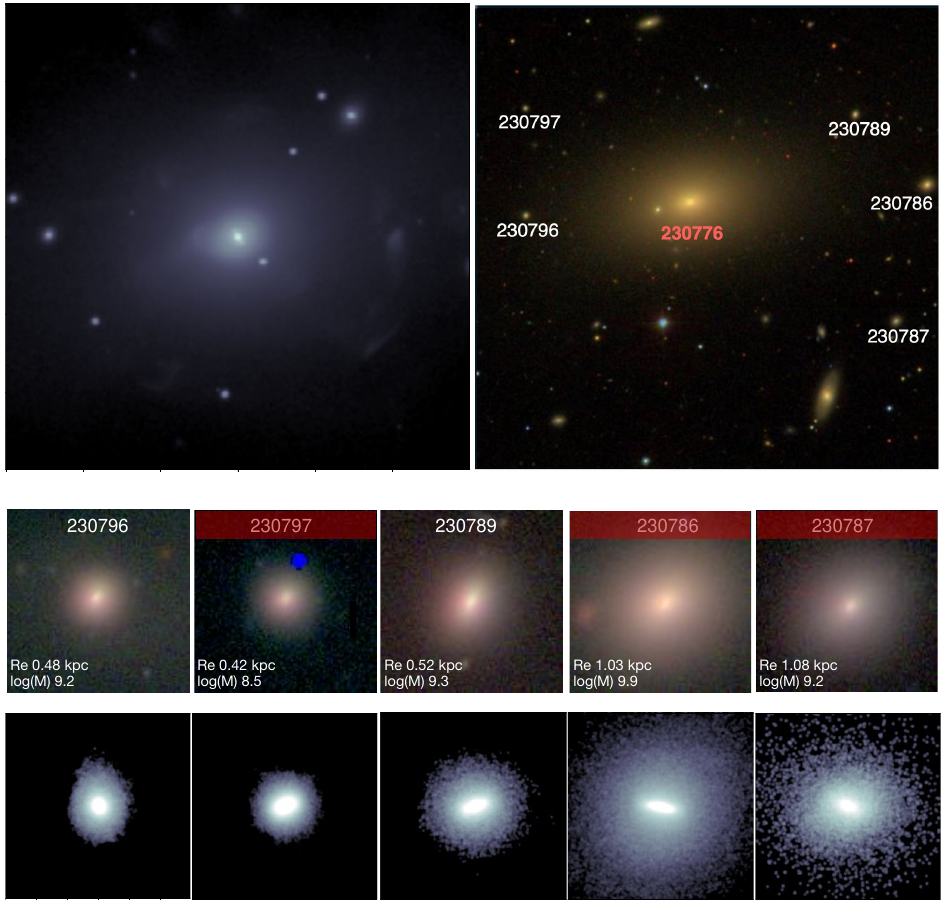}
   \caption{Systems with multiple cEs around a single host galaxy in the simulation (top left) and from the observational sample (top right). Three of the galaxies in the observed system are marginally outside the definition of a cE used here, however as noted previously the strict definition is somewhat arbitrary and we retains these additional galaxies being compact high-mass systems. The bottom two rows show images of individual cEs in the observed (top, HSC images) and simulated (bottom) system, arranged from left to right by their distance from the host. The observed galaxies falling outside the cE definition are highlighted by red shading across their SAMI ID numbers, with their effective radii and mass included to show where they lie in relation to the cE limit. }
 \label{multi_ce}
 \end{figure*}

\subsubsection{Systems of cEs around a single host}

In the SAMI observational sample, we identified a system containing three cEs near a single host galaxy, along with two more galaxies which are marginally above the 1 kpc threshold of being classed as a cE (and which were below this threshold in a previous version of the GAMA catalogue). In addition, we also found a second system of two cEs around a single host. We note that this likely underestimates the number of multiple systems, and the number of cEs per system, since the target selection of SAMI, particularly for these smaller galaxies is not complete. 

We looked for systems of multiple cEs orbiting around a single host within the simulation, to further understand how common such systems are and how they evolve. We found that out of 71 large galaxies found to host at least 1 cE, 30 hosted more than 1 cE. 14 hosted 2 cEs, 9 hosted 3, 4 hosted 4 and 3 hosted 5 or more. The largest system contains 15 cEs. One of the larger systems is illustrated in Figure~\ref{multi_ce}, along with the multiple system observed in SAMI. 

\begin{figure*}
\includegraphics[width=1.9\columnwidth]{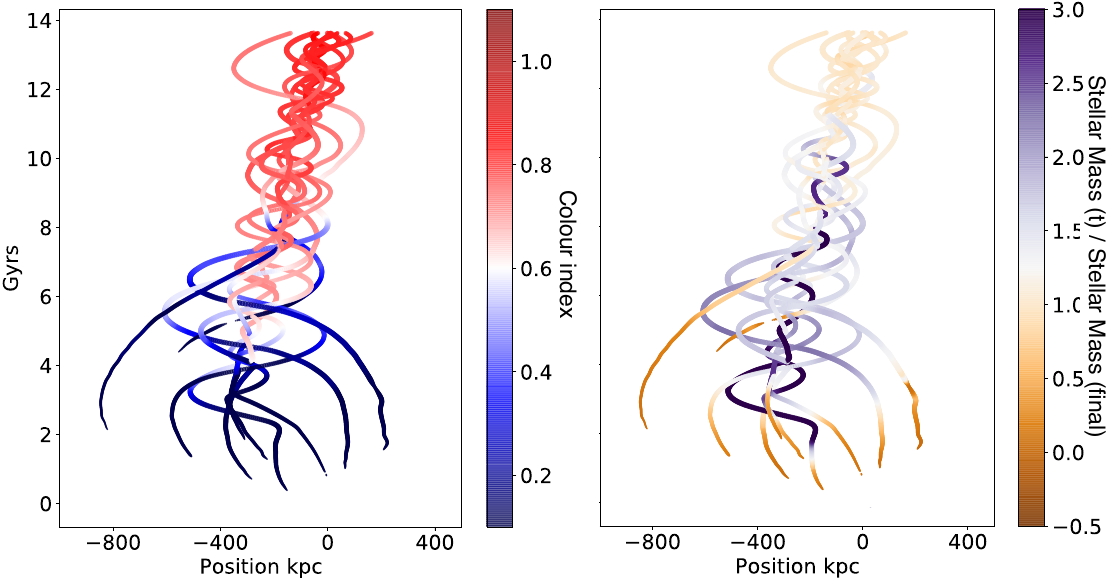}
   \caption{Position-time plots illustrating the evolution of the system of cEs shown in Figure~\ref{multi_ce}. The plots display the individual tracks of all galaxies which end up as a cE within 500 kpc of the central host galaxy, projected onto the x-axis (with points layered according to their y-axis positions). The left plot shows the colour evolution of the galaxies while the right plot shows the stellar mass evolution in relation to the galaxy's final mass. All galaxies become red and passive by around 8 Gyrs and lose the majority of their stellar mass by 11 Gyrs. Note the galaxy forming via the growth + infall pathway, seen as the galaxy coming in from the far left and remaining below its final mass (brown) for the majority of the time. Other projections confirm that this is the galaxy seen to maintain a wider orbit despite a slight interaction at ~8 Gyrs.}
 \label{system_ev}
 \end{figure*}

In the simulated system, three of the 15 cEs formed through the growth + infall pathway with the remaining forming via the stripped pathway (including the example galaxy shown in Figure~\ref{stripped}). The full evolution of the system is displayed in Figure~\ref{system_ev}, with the colour evolution of the eventual cE galaxies shown on the left and their stellar mass evolution shown on the right. The progenitor galaxies begin in different locations and converge independently on the host, showing that the multiple system did not arise due to a single event such as the infall of a galaxy group. Most of these galaxies, after initially building up a significant amount of mass (indicated by the purple colours in the right-hand plot) enter into tight orbits around the host by around 8 Gyrs, and gradually lose their mass due to tidal forces from the host (indicated by the transition to brown colours in the plot). The exception to this is the galaxy which can be seen coming in from the far left - rather than building up a large mass which is subsequently stripped, this galaxy always has a small mass and is a cE forming via the growth + infall pathway. Unlike the others, it maintains a wider orbit around the host, seen as the looping path first to the right then to the left as time approaches the present day.

\subsubsection{Embedded Disks}

Around 11 percent of the compact galaxies in the simulation show signs of embedded disks in the final snapshot. All of these galaxies formed via the growth formation pathway. The disks are the sites of ongoing star formation activity and are subsequently most visible in the g-band luminosity images. Such disks also appear during the evolution of many other galaxies in this group, as well as those which then subsequently fall into a galaxy group. These disks appear to form out of infalling gas which collapses into a disk structure. The disks become smaller as the gas concentrates towards the core, and eventually disappear once the infall of fresh gas ceases.

As described earlier, observed cEs in our observational sample also display evidence of embedded disk structures, more so in isolation than amongst those near a host. The fact that we see such structures within simulated galaxies originating through the growth pathway (but not via the stripped pathway) suggests that such galaxies are likely to have formed via this pathway and are not stripped remnants of larger galaxies. In the case of host-associated cEs having embedded disks, these may be of the growth+infall pathway, where recent star formation took place within the disk structure before the galaxy fell into the vicinity of its host system and lost its star forming gas. Due to the recent concentration of star formation within the disk, the disk-like structure in the stellar distribution may still be visible after the gas has been stripped and the galaxy has evolved towards redder colours. Such evolutionary paths can be found within our simulation sample, and indeed the third galaxy shown in the top row of Figure~\ref{disks} (with a clearly visible disk) is one such case (see the figure in the appendix for this galaxy's complete evolution). Since these disk structures would not be expected to be seen in cEs formed via stripping, they may be an additional diagnostic to identify galaxies forming via the growth+infall process.

\subsubsection{Host vs Isolated cEs}

Next we looked to confirm if the different formation pathways illuminated above lead to the compact ellipticals observed in different environments, namely those near a host galaxy and those in isolation. For each cE in our sample, we searched through the locations of all galaxies in the final snapshot with stellar masses greater than $10^{10.5}M_{\odot}$ (as was done for the observation sample) and located the nearest such galaxy. The resulting relationship between the formation pathway and the distance to the nearest potential host galaxy is shown in Figure~\ref{host}. We find that the majority of stripped cEs are located within 500 kpc of a large galaxy. We investigated the stripped cEs located at greater distances using space-time diagrams such as those in Figure~\ref{stripped} and found that these galaxies are held in tight orbits around host galaxies with masses below $10^{10.5}M_{\odot}$. Therefore, it appears that all stripped cEs remain close to the galaxy which stripped them. This is in contrast to the proposal put forward by \citet{2015Sci...348..418C} to explain isolated cEs, where they suggested that these cEs may be stripped nuclei which were subsequently thrown out of the cluster. However, we note again that the TNG50 simulation does not contain large cluster environments, and so therefore we cannot rule out that this occurs in higher density environments. 

Meanwhile, the cEs formed via continuous growth form a distinct population at greater distances from larger galaxies. These two populations are linked together by the third growth+infall group, where the compact galaxies formed in isolation have subsequently fallen into a galaxy group. However, while these galaxies are unsurprisingly found closer to larger galaxies than those which have remained isolated, their distances from the large galaxy cover a broader range than those formed via the stripped pathway; in particular, very few of these galaxies are found within 100 kpc of a large galaxy. Figure~\ref{host} also highlights the relation between colour and distance from the closest large galaxy, with host-associated cEs being significantly redder than those further out. The spatial gap seen in log space between the host-associated and isolated populations is consistent with the binomial population recently reported in \citet{2022ApJ...934L..35C}.

\subsubsection{Comparison with Observations}

After establishing that the stripped and growth pathways of cE formation lead to host-associated and isolated cEs respectively, we next look to see if we can use these two pathways to reproduce the observed differences between the two populations; namely the higher metallicity, redder colours and smaller effective radii seen in the host-associated population relative to isolated cEs. 

\begin{figure*}
\includegraphics[width=1.8\columnwidth]{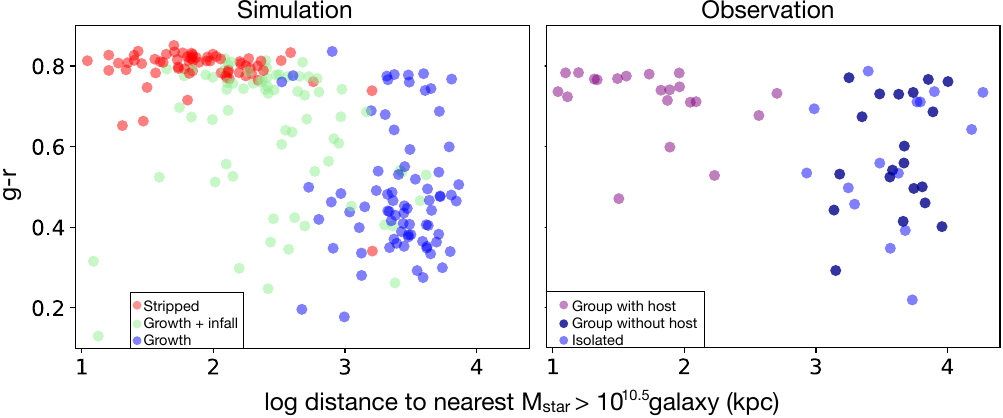}
   \caption{Relation between the cEs and their nearest galaxy with a stellar mass greater than $10^{10.5}M_{\odot}$. The left panel shows the relation between the colour index and the distance to the nearest large galaxy, for the simulation sample, with stripped cEs shown in red, growth cEs in blue and growth+infall cEs in green. This clearly shows that most cEs located next to a large galaxy formed via stripping while those in isolation formed via the growth pathway. The growth+infall cEs bridge these two populations, being located near large galaxies but covering a broader range of radii. The right panel shows the same plot for the observations, with purple points being cEs located in a group with a galaxy > $10^{10.5}M_{\odot}$, dark blue points showing cEs in groups with only low-mass galaxies and light blue points being galaxies outside of any group. The distribution of cEs in the simulation is very similar to the observations.}
 \label{host}
 \end{figure*}

\begin{figure}
\includegraphics[width=1\columnwidth]{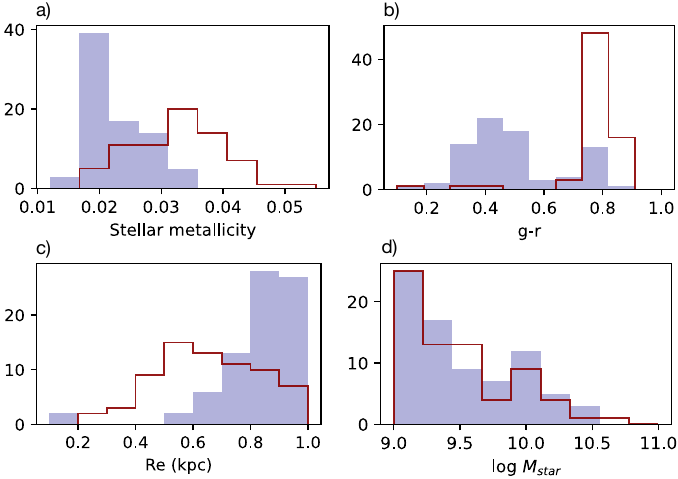}
   \caption{Distributions of observable properties between cEs originating via the stripped pathway (red) and those originating via the continuous growth pathway (blue) for the simulated sample. a) shows the metallicity (here defined as the mass in metals divided by the total mass), b) the colour index, c) the half-mass radius and d) the stellar mass. The stripped cEs are seen to feature smaller radii, redder colours and higher metallicities relative to the growth cEs, as has been observed for host-associated cEs relative to those in isolation. Meanwhile the stellar mass distributions are similar, showing that the two pathways produce galaxies of similar mass.}
 \label{obs_comp}
 \end{figure}

Figure~\ref{obs_comp} b) shows the colour distribution of the final cEs. As was also clearly evident from Figure~\ref{host}, the stripped cEs are all very red while the mass-growth cEs have a wide range of colours, with some being very blue. This is in line with the difference between the observed isolated and host-associated cEs in Figure~\ref{distribution}. The metallicity distributions are shown in Figure~\ref{obs_comp} c); those forming via the stripping pathway are significantly more metal-rich than the in-situ cE population, with very little overlap between them. This is again consistent with the observed difference in metallicity between isolated and host-associated cEs, with the former having lower metallicities than the latter. These relative distributions are also present in the compact galaxies with $10^{8} < M/M_{\odot} < 10^{9}$ (Figure~\ref{low_mass} e-h in the Appendix), with the exception that the number of galaxies forming through the growth pathway increases with decreasing mass. 

Finally, we investigate the kinematic differences between the host-associated and isolated cEs in the observational sample, and look to try and reproduce this from the formation pathways seen in the simulations. Figure~\ref{kinematics} displays the radial kinematic profiles for the observed (a-d) and simulated (f-i) galaxies, separated into host or isolated cEs in the case of the observations, and by the growth and stripped pathways in the simulation.

For the observed samples, all galaxies tend to be dispersion dominated, as indicated by the $v/\sigma$ distribution in Figure~\ref{kinematics} e). The isolated cEs tend towards higher degrees of rotational support relative to the host-associated cEs. This is also evident in the radial plots, where we see the velocity profiles of the host-associated cEs all remain low while the range in velocities of the isolated galaxies spread up to higher velocities. The range of these velocity profiles, as well as the differences between the two populations are reflected in the simulation sample, when we separate them into those forming through the stripped pathway and those formed through the growth pathway. This grouping also reproduces the difference in $v/\sigma$ between the two populations, with stripped cEs having lower degrees of rotational support compared to the growth cEs. Interestingly the cE galaxies located near a host galaxy have similar degrees of rotational support as the high-mass elliptical galaxies known as 'slow rotators' \citep{2019A&A...632A..59F}. This may be caused by dynamical heating of the cE during close passages to the host - for example, the dwarf galaxy in a close orbit around a host simulated by \citet{2019ApJ...875...58D} experienced progressively decreasing $v/\sigma$ as it orbited (note however that this galaxy started the simulation as an extended dwarf rather than as a compact galaxy). The isolated cEs, meanwhile, have similar $v/\sigma$ values to more extended dwarf elliptical galaxies with masses between $10^{8}$ and $10^{9.5} M_{\odot}$ \citep{2020MNRAS.497.1571S}. Dwarf elliptical galaxies may also have formed primordially rather than being stripped remnants  \citep{2009MNRAS.396.2133K}, or the harassment of a late-type galaxy \citep{2009A&A...494..891A}

However, interestingly the $v/\sigma$ distribution of growth-originating cEs includes galaxies with higher degrees of rotational support compared to the observations. This may in part be due to the selection affects of the SAMI sample - we may be missing highly isolated systems in the observational sample which may have higher rotational support. Alternatively, it could be suggestive of an important difference in the kinematics of the simulated galaxies relative to observations, which may need to be further investigated in future work.

\begin{figure*}
\includegraphics[width=2\columnwidth]{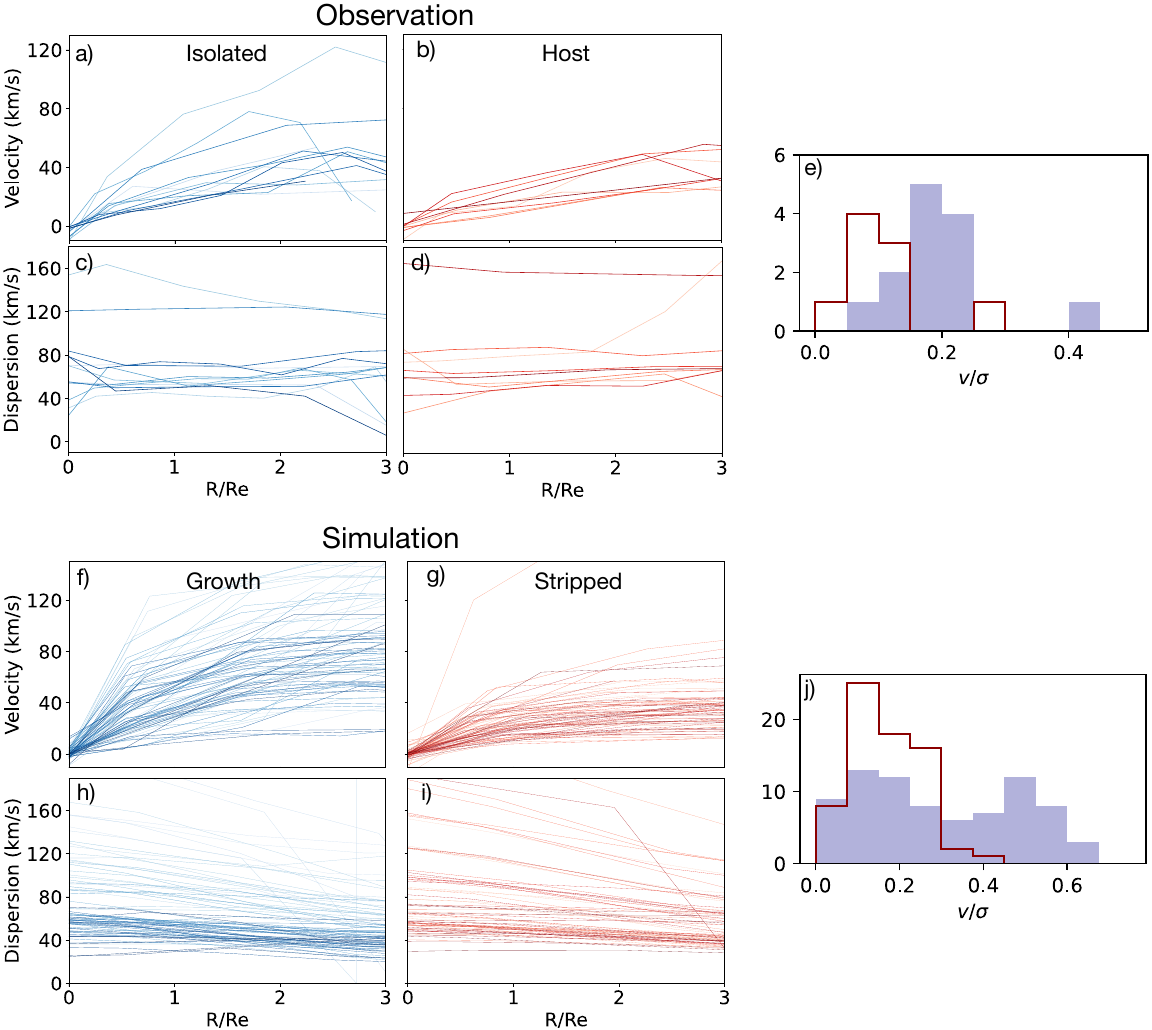}
   \caption{Comparison of the stellar kinematics of cE galaxies in the observation and simulated sample. The observed sample (a-e) is separated into host associated (red) and isolated (blue) galaxies, while the simulation sample (f-j) is separated into those forming via the stripped (red) or growth (blue) pathway. a-b and f-g show the line-of-sight velocity profiles, while c-d and h-i show the velocity dispersion profiles. The histograms on the right show the distribution in $v/\sigma$ for the observations (e) and simulation (j). In the observed sample, isolated galaxies tend towards higher line-of-sight velocities and higher degrees of rotational support relative to the host-associated sample. This difference is reflected in the simulations when separated by formation pathway, indicating the formation pathway can explain the observed kinematic differences.}
 \label{kinematics}
 \label{kinematics}
 \end{figure*}

\section{Discussion}

\subsection{Two pathways}

Recent observation have suggested that there are two environment-dependent populations of compact elliptical galaxies. After analysing the mass evolution of the compact elliptical galaxies in IllustrisTNG-50, we see that these galaxies can indeed arise from two completely different processes, namely that of stripping of a larger disk galaxy and by continuous growth. 

\subsection{Isolated cEs}

The observed differences we found between the host and isolated cE populations agree with previous findings, in particular the isolated cEs are younger, bluer and more extended than their isolated counterparts \citep{2018MNRAS.473.1819F}. This adds further observational support to the idea that isolated cE galaxies form through a distinct pathway, and indeed this is what we found within the simulation.

Around $68\pm4$ percent of all compact galaxies in the simulation formed intrinsically through continuous star formation, with $36\pm5$ percent remaining outside of group environments in isolation. These galaxies formed out of collapsing gas clouds, with very few (if any) merger events. Initially, the star formation is distributed over a large area, creating a diffuse dwarf galaxy. The gas and accompanying star formation then becomes concentrated in the centre, where it gradually builds up the stellar density within the galaxy core. Most of these galaxies remain blue and star forming up until the present day, since there are no external influences disturbing their surrounding gas halo. Whether all of these should be placed in the same class as the traditional cE galaxies is uncertain, since cE galaxies are normally taken to be passive. However, the majority of these galaxies exhibit similar morphologies, and those which have become passive resemble classical cE galaxies. These star-forming galaxies resemble previously observed compact star forming galaxies in mass and radius \citep{2015ApJ...807..139F}, and therefore the findings here suggest such galaxies will ultimately evolve into isolated cE galaxies. 

These same pathways are also present in the population with stellar masses in the range $10^{8} < M_{\odot} < 10^{9}$, however there are relatively far more compact galaxies forming through the growth pathway compared to the stripped pathway. Note that the low-mass sample included here is a sub-sample, while our sample of galaxies above $10^{9} M_{\odot}$ is complete for the simulation. When re-scaled to account for the sub-sampling of the lower-mass sample, the absolute number of lower-mass galaxies forming via the stripped pathway is higher than the number of stripped galaxies with final masses above $10^{9} M_{\odot}$. The difference in the relative numbers of galaxies forming through each pathway is therefore a reflection of there being far more lower-mass galaxies forming in isolation through the growth pathway, suggesting that it is less common for a high-mass compact galaxy to be able to form in isolation.

Importantly, we didn't find any contribution from galaxies ejected out of galaxy groups to the isolated population, in contrast to the expectations in \citet{2015Sci...348..418C}. All isolated cE galaxies formed in isolation, through a pathway distinct to that of the majority of host-associated cE galaxies (see below). 

\subsection{Host-associated cEs}

Around $32\pm6$ percent of compact elliptical galaxies were found to form via the stripped pathway, all of which were still associated with their host galaxy in the final snapshot. These galaxies initially grow into a large, typical spiral galaxy with significant amounts of star formation activity. These then fall into a larger galaxy group and enter an orbit about the group's central galaxy. As they do so, a significant amount of gas is stripped out of the galaxy, bringing an end to star formation activity, and the spiral galaxy arms fade to leave a smooth, passive disk reminiscent of an S0. As the orbit becomes tighter, the outer disk is then stripped away to leave behind the central nucleus of the original galaxy, which takes the form of a compact elliptical. This follows the model of \citep{2001ApJ...557L..39B} for the formation of M32 and shows that it is common in a cosmological context. These findings also show that the progenitors are ubiquitously spiral galaxies rather than elliptical galaxies or a combination of both. While globular clusters are below the resolution limit of the simulation, the number of cE galaxies formed from the stripping pathway would suggest the merging star cluster pathway suggested by \citep{2019MNRAS.489.2746U} is unlikely. 

The stripped formation pathway has a very similar appearance to the S0 galaxies forming via cluster infalls as studied in \citet{2021MNRAS.508..895D}, with the main difference being that in this case, the galaxy gets much closer to the central galaxy, and hence experiences tidal forces strong enough to strip away the stellar disk. As mentioned above, these galaxies indeed quickly go through an S0 phase, and some of the in-falling S0 galaxies identified in the previous work were also seen to have shrinking disks, perhaps being in the early stages of having their outer disk stripped away completely. Ultra-compact dwarf galaxies may also be the stripped nuclei \citep{2021MNRAS.501.1852M} or stripped nuclear star clusters \citep{2020A&ARv..28....4N} of larger galaxies, and perhaps cE galaxies closer to the host will be stripped further to form such galaxies. Large ultra-compact dwarfs, cE and S0 galaxies around a larger host galaxy may therefore represent a continuous process of stellar stripping, with the galaxies getting more compact as their distance from the host decreases.  

Being the remnants of the bulges of spiral galaxies, which are known to have older stellar populations and high metallicities, we expect that these galaxies will be red and metal-rich, and indeed that is what we found. The higher metallicities are consistent with observations \citep{2020ApJ...903...65K}, showing that the higher metallicities are due to these galaxies originating from more massive spiral galaxies. In addition, the effective radii of these galaxies are smaller relative to the rest of the cE galaxy population, also matching observations \citep{2015ApJ...807..139F}. All of these findings also agree with observations of cE galaxies near a potential host galaxy within this study, and indicate that most of these galaxies are not relic galaxies but are instead the remnant bulges of massive spiral galaxies. 

In addition, some cE galaxies formed in isolation and subsequently fell into a galaxy group or cluster. While most of these galaxies remain on wider orbits, a small number do fall into the near vicinity of a more massive galaxy. Therefore, while the majority of host-associated cE galaxies are stripped spiral galaxies, there is also a small population of galaxies in the same environment which formed intrinsically. These galaxies may be able to be separated by looking at their central black hole mass \citep{2021ApJ...917L...9R}; since the cEs forming through the stripped pathway were originally a much more massive galaxy, they would be expected to harbour a more massive central black hole. 

\subsection{Embedded disks}

We found a number of cE galaxies with apparent embedded disks evident in the optical imagery, both in the field and near a host. These appear to be blue star-forming disks in the isolated galaxies and passive disks in the host-associated galaxies. Interestingly, within the simulation we also see such disks; around 11 percent of isolated galaxies in the simulation feature embedded disk structures, which are most prominent in g-band imagery. These also usually coincide with the location of ongoing star formation and are most visible in the g-band flux. This raises the question of whether host-associated cE galaxies with embedded disks may have originally formed in isolation and then fell into the vicinity of a host; such disks may therefore be a way to identify cE galaxies around a host which originated from the different formation pathways.  Two such galaxies are seen in the observational sample; these cE galaxies are close to a large host galaxy and both feature a clear edge-on disk structure embedded within their stellar halos. However, nuclear disks have also been observed in S0 galaxies \citep{2002ApJ...573..131P}, so this alone may not be a conclusive indication of how they formed. 

While disks have been observed in more extended dwarf elliptical galaxies previously \citep{2003A&A...400..119D}, in some cases even with spiral structures \citep{2000A&A...358..845J}, such disks haven't  been observed in cE galaxies before. Given the small size of these galaxies and the embedded nature of their disks, these are likely very difficult to discern in observations, and it is only with a large sample of relatively high-resolution imagery that they became evident.

\section{Summary and conclusions}

In this paper we have compared matched samples of observed and simulated  compact elliptical (cE) galaxies to  determine how they form. We defined cE galaxies by the same mass (stellar mass greater than $10^{9} M_{\odot}$) and size (effective radii less than 1 kpc) criteria in both samples. We did not apply any colour selection.

\begin{enumerate}
\item We selected our observational sample of cE galaxies from the SAMI galaxy survey according to the criteria above. We found 56 such galaxies. 20 cE galaxies resided within 500 kpc of a high-mass host galaxy, with the remainder being in isolation. The cE galaxies near a host feature smaller radii, redder colours, lower star formation rates and a lower degree of rotational support relative to the isolated population, which is suggestive of different formation pathways for cE galaxies in different environments. 

\item We searched the IllustrisTNG-50 cosmological simulation for cE galaxies, using the same selection criteria as above, and found 234 such galaxies. Following these galaxies back through time, we found that they form through two main formation pathways. In the first pathway (comprising $32\pm5$ percent of the cE population), the galaxy initially grows into a large spiral galaxy, which then falls into a tight orbit around a higher-mass host galaxy. Initially the gas is stripped out of the galaxy, then as the orbit becomes tighter, the outer stellar disk is stripped away by tidal forces, leaving behind the compact core. The remaining $68\pm4$ percent of cE galaxies form as compact galaxies in isolation, and experience a continuous increase in mass throughout their evolution, with no significant stripping events.

\item All the simulated galaxies that form by the stripped pathway remain close to their host galaxy, while all those in isolation formed through the continuous-growth pathway. We therefore looked to explain the observed differences in cE galaxies near a host or in isolation using these two formation pathways. We found that relative to those forming in isolation, cE galaxies forming through the stripped pathway have smaller radii, redder colors and a lower degree of rotational support, in agreement with the observed cE galaxies near a high-mass host. 

\end{enumerate}

We therefore conclude that cE galaxies form by two different pathways: the stripping of larger spiral galaxies in dense environments and the continuous growth of isolated galaxies. The observed differences in cE galaxy properties with environment are explained by the relative importance of the two pathways in different environments.

\section*{Acknowledgements}

The SAMI Galaxy Survey is based on observations made at the Anglo-Australian Telescope. The SAMI was developed jointly by the University of Sydney and the Australian Astronomical Observatory. The SAMI input catalogue is based on data taken from the Sloan Digital Sky Survey, the GAMA Survey, and the VST ATLAS Survey. The SAMI Galaxy Survey is supported by the Australian Research Council centre of Excellence for All Sky Astrophysics in 3 Dimensions (ASTRO 3D), through project number ce170100013, the Australian Research Council centre of Excellence for All-sky Astrophysics (CAASTRO), through project number ce110001020, and other participating institutions. The SAMI Galaxy Survey website is http://sami-survey.org/.
 
GAMA is a joint European-Australasian project based around a spectroscopic campaign using the Anglo-Australian Telescope. The GAMA input catalogue is based on data taken from the Sloan Digital Sky Survey and the UKIRT Infrared Deep Sky Survey. Complementary imaging of the GAMA regions is being obtained by a number of independent survey programmes including GALEX MIS, VST KiDS, VISTA VIKING, WISE, Herschel ATLAS, GMRT, and ASKAP providing ultraviolet to radio coverage. GAMA is funded by the STFC (UK), the ARC (Australia), the AAO, and the participating institutions. The GAMA website is http://www.gama-survey.org/
 
This work was supported through the Australian Research Council's Discovery Projects funding scheme (DP170102344). SMS acknowledges funding from the Australian Research Council (DE220100003).

\section*{Data Availability} 
The observational data used in this article was accessed from the publicly available SAMI Data Release 3 \citep{2021MNRAS.505..991C}, available through https://datacentral.org.au. The simulation data underlying this article were accessed from the publicly available IllustrisTNG-50 simulation, available at https://www.tng-project.org/data. The derived data generated in this research will be shared on reasonable request to the corresponding author.



\bibliographystyle{mnras}



\appendix

\begin{figure*}
\includegraphics[width=2\columnwidth]{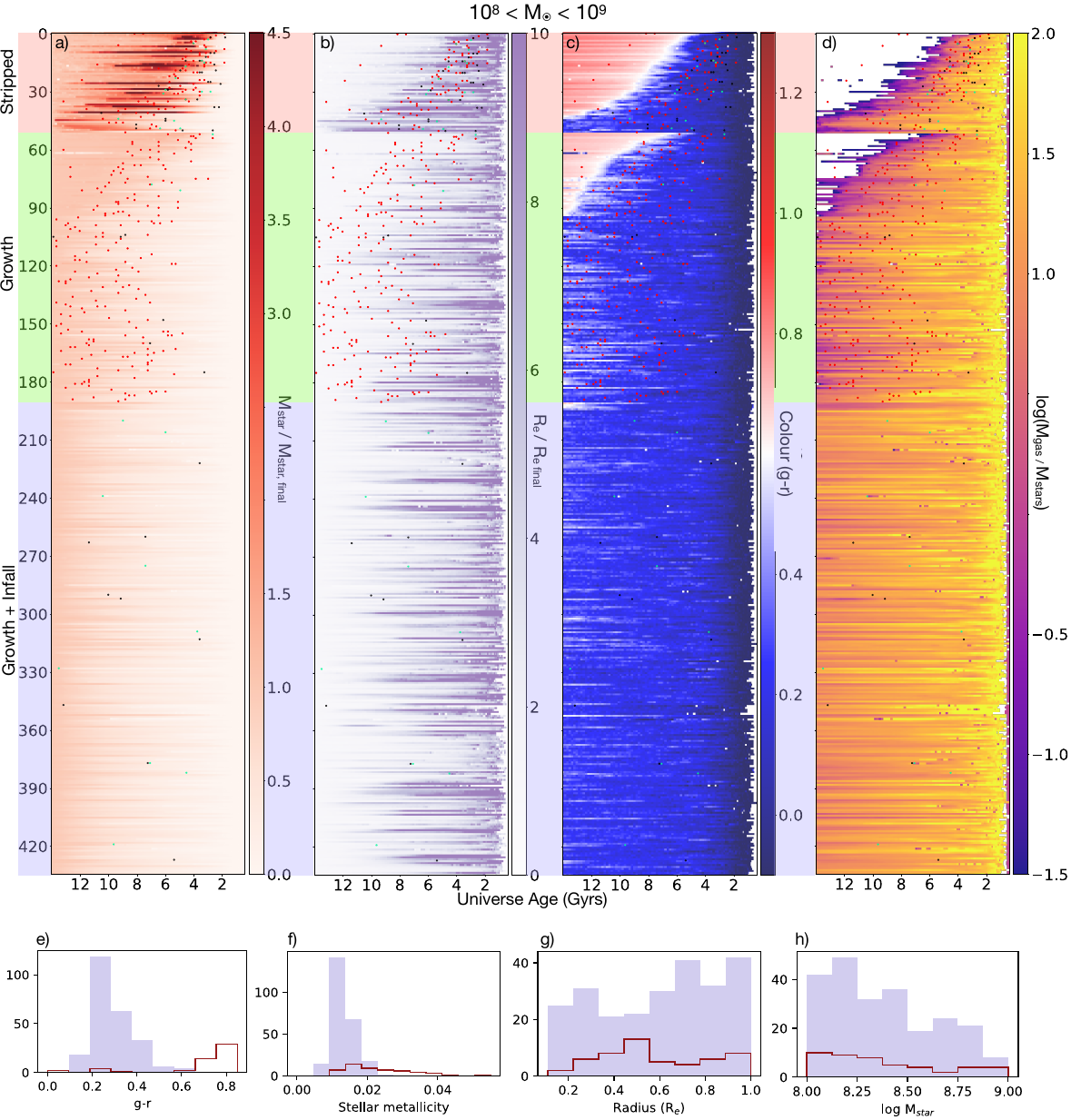}
   \caption{Evolution and properties of lower-mass compact galaxies with stellar masses in the range $10^{8} < M_{\odot} < 10^{10}$. a) displays the evolution of mass relative to the galaxy's final mass, separated into the same groups as in Figure~\ref{mass_evolution}. b) shows the evolution in radius relative to the final radius, c) shows the colour evolution and d) shows the gas fraction evolution. Within this mass range, a smaller portion of the population originates through the stripping of a larger galaxy, in contrast to the compact galaxies with masses above $10^{9} M_{\odot}$. e) shows the distributions of colour for galaxies forming through the stripped (red) and growth (blue) pathway, f) shows the stellar metallicity distributions, g) the radius distributions and h) the mass distributions. The relative distributions show the same trends as for the higher-mass compact galaxies, with the exception of mass, where the relative number of galaxies forming through the continuous growth pathway increases as the mass decreases.}
\label{low_mass}
 \end{figure*}


\bsp	
\label{lastpage}
\end{document}